\documentclass[aps,prl,showpacs,twocolumn]{revtex4}
\usepackage{amsfonts}
\usepackage{amssymb}
\usepackage{amsmath}
\usepackage{graphicx}
\usepackage{epsfig}

\begin{document}

\title{Probing Mott lobes via the AC Josephson effect}
\author{M.X. Huo$^{1}$}
\author{Ying Li$^{1}$}
\author{Z. Song$^{1}$}
\email{songtc@nankai.edu.cn}
\author{C.P. Sun$^{2}$}
\email{suncp@itp.ac.cn}
\homepage{http://www.itp.ac.cn/~suncp}
\affiliation{$^{1}$Department of Physics, Nankai University, Tianjin 300071, China}
\affiliation{$^{2}$Institute of Theoretical Physics, Chinese Academy of Sciences,
Beijing, 100080, China}

\begin{abstract}
The alternating-current (AC) Josephson effect is studied in a system
consisting of two weakly coupled Bose Hubbard models. In the framework of
the mean field theory, Gross-Pitaevskii equations show that the amplitude of
the Josephson current is proportional to the product of superfluid order
parameters. In addition, the chemical potential--current relation for a
small size system is obtained via the exact numerical computation. This
allows us to propose a feasible experimental scheme to measure the Mott
lobes of the quantum phase transition.
\end{abstract}

\maketitle

\emph{Introduction.} Recent development of experiments allows the detection
of the quantum phase transition in strongly correlated many-body systems
\cite{Probe}. The relevant physics is captured by the Bose-Hubbard model,
which describes the competition between the kinetic-energy and
potential-energy effects. The Mott insulator (MI) to superfluid (SF) phase
transition of the Bose-Hubbard model was described qualitatively using a
mean field theory \cite{Fisher} and realized in a gas of ultracold atoms
\cite{QBEC}. Generally, for the MI to SF transition, the superfluid order
parameter predicted by the mean field theory can not be measured directly.
However, in this letter, we propose a feasible realization to detect the
phase transition and obtain the Mott lobes of the order parameter. The main
idea relates to the well known alternating-current Josephson effect, which
has been well investigated in the Bose-Einstein condensates (BEC) \cite{Jc,
Jc2, Jc3, Jc4}.

The Josephson junction is composed of two superfluid parts separated by a
tunnelling barrier. When a constant chemical potential difference is
applied, the AC Josephson effect occurs, correponding an oscillating
particle flow through the barrier. If one part is fixed in surperfluid
state, the phase of other part should determine the amplitude of the AC
current. Inversely, it is possible to detect the state of the target part
via measuring the Josephson current across the barrier.

In this letter, through the well established mean field theory and
spontaneous symmetry breaking mechanism for quantum phase transition, the
analytical analysis reveals that the amplitude of the Josephson current is
proportional to the product of superfluid order parameters of the two parts.
Through the measurement of the AC Josephson current, the Mott lobes can be
obtained indirectly. Moreover, numerical simulations show that the profile
of the Mott lobes for small size system is consistent with that in
thermodynamic limit. This opens a possibility to detect the BEC in small
quantum device. As an application we discuss an experimental realization in
a cavity-atom hybrid system based on the recent work in Ref. \cite%
{Photonnonlinearity}.


\begin{figure}[tbp]
\includegraphics[bb=77 569 527 778, width=7 cm]{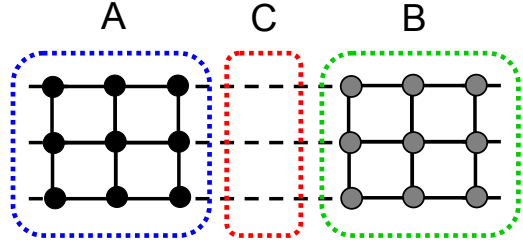}
\caption{\textit{(Color online) Schematic plot of the lattice model used in
this work. (A) is a Bose-Hubbard model with the on site repulsion }$U$%
\textit{\ triggering the Mott transition; (B) is a free boson lattice model
with a chemical potential }$\protect\mu $\textit{\ which controls the
density of bosons in lattice A; (C) is a thin contact surface between A and
B which can be represented by a direct weak tunnelling across A and B or a
free boson lattice model with an on site repulsion larger than }$\protect\mu
$\textit{.}}
\label{Model}
\end{figure}


\emph{Hamiltonian and the mean field theory. }The setup is depicted as the
following Hamiltonian
\begin{equation}
H=H_{A}+H_{B}+H_{C},  \label{H1}
\end{equation}%
where
\begin{eqnarray}
H_{A} &=&-\kappa \sum_{\left\langle \mathbf{i},\mathbf{j}\right\rangle \in
A}\left( a_{\mathbf{i}}^{\dag }a_{\mathbf{j}}+h.c.\right) +\frac{U}{2}\sum_{%
\mathbf{i}\in A}a_{\mathbf{i}}^{\dag }a_{\mathbf{i}}^{\dag }a_{\mathbf{i}}a_{%
\mathbf{i}},  \notag \\
H_{B} &=&-\kappa \sum_{\left\langle \mathbf{i},\mathbf{j}\right\rangle \in
B}\left( b_{\mathbf{i}}^{\dag }b_{\mathbf{j}}+h.c.\right) +\mu \sum_{\mathbf{%
i}\in B}b_{\mathbf{i}}^{\dag }b_{\mathbf{i}},  \label{Habc} \\
H_{C} &=&-g\sum_{\mathbf{i}\in C}\left( a_{\mathbf{i}}^{\dag }b_{\mathbf{i}%
}+h.c.\right) .  \notag
\end{eqnarray}%
Here $a_{\mathbf{i}}$ and $b_{\mathbf{i}}$ are the boson operators. In this
paper, we consider lattices A and B having an identical structure with size $%
N$ for simplicity. Each lattice site $\mathbf{i}$ $(=1,2,...,N)$ corresponds
to two positions in lattices A and B, respectively. Among them, $\mathbf{i}%
\in C $\ lies on the contact surface. We consider the weakly coupled case $%
g\ll \kappa $. Fig. 1 is a schematic representation of the setup.

We start our investigation from the mean field approximation. The order
parameters can be defined as the expectation values of boson operators $a_{%
\mathbf{i}}(t)$ and $b_{\mathbf{i}}(t)$ in the Heisenberg picture by
ignoring the fluctuations, i.e.,
\begin{equation}
\psi _{a,\mathbf{i}}(t)\equiv \left\langle a_{\mathbf{i}}(t)\right\rangle ,%
\text{ }\psi _{b,\mathbf{i}}(t)\equiv \left\langle b_{\mathbf{i}%
}(t)\right\rangle ,  \label{OP}
\end{equation}%
where the average is taken with respect to the ground state. In this paper,
we use $a_{\mathbf{i}}$ and $b_{\mathbf{i}}$ to denote the boson operators
in the Schr$\ddot{o}$dinger picture. In order to study the time evolution of
the order parameter, we make use of the Gross-Pitaevskii (GP) equations in
the lattice \cite{GPE, GPE2}. The GP equations corresponding to the
Hamiltonian (\ref{H1}) read
\begin{subequations}
\begin{eqnarray}
i\frac{\partial }{\partial t}\psi _{a,\mathbf{i}} &=&-\kappa
\sum_{\left\langle \mathbf{i},\mathbf{j}\right\rangle }\psi _{a,\mathbf{j}%
}+U\left\vert \psi _{a,\mathbf{i}}\right\vert ^{2}\psi _{a,\mathbf{i}}-g_{%
\mathbf{i}}\psi _{b,\mathbf{i}},  \label{GPa} \\
i\frac{\partial }{\partial t}\psi _{b,\mathbf{i}} &=&-\kappa
\sum_{\left\langle \mathbf{i},\mathbf{j}\right\rangle }\psi _{b,\mathbf{j}%
}+\mu \psi _{b,\mathbf{i}}-g_{\mathbf{i}}\psi _{a,\mathbf{i}},  \label{GPb}
\end{eqnarray}%
where $g_{\mathbf{i}}=g$ for $\mathbf{i}\in C$ and $g_{\mathbf{i}}=0$ for $%
\mathbf{i}\notin C$.

When $g$ is small enough, we can define solutions of Eqs. (\ref{GPa}) and (%
\ref{GPb}) as
\end{subequations}
\begin{equation}
\psi _{a,\mathbf{i}}=\phi _{a}\left( t\right) \Phi _{a,\mathbf{i}},\text{ }%
\psi _{b,\mathbf{i}}=\phi _{b}\left( t\right) \Phi _{b,\mathbf{i}},
\label{Psi_ai}
\end{equation}%
where $\Phi _{a,\mathbf{i}}$ and $\Phi _{b,\mathbf{i}}$ are solutions for
the ground states of Eqs. (\ref{GPa}) and (\ref{GPb}) when $g=0$. $\Phi _{a,%
\mathbf{i}}$ and $\Phi _{b,\mathbf{i}}$ are set to be real and normalized as
$\sum_{\mathbf{i}}\Phi _{a,\mathbf{i}}^{2}=\sum_{\mathbf{i}}\Phi _{b,\mathbf{%
i}}^{2}=1.$ Taking the periodic boundary condition, $\Phi _{a,\mathbf{i}}=$ $%
\Phi _{b,\mathbf{i}}$ $=1/\sqrt{N}$. After replacing $\psi_{a,\mathbf{i}}$
and $\psi_{b,\mathbf{i}}$, GP equations (\ref{GPa}) and (\ref{GPb}) become
the two-state model \cite{Jcformula}
\begin{subequations}
\begin{eqnarray}
i\frac{\partial }{\partial t}\phi _{a}\left( t\right) &=&\left[
E_{a}+U_{a}\left\vert \phi _{a}\left( t\right) \right\vert ^{2}\right] \phi
_{a}\left( t\right) -K\phi _{b}\left( t\right),  \label{TSM_a} \\
i\frac{\partial }{\partial t}\phi _{b}\left( t\right) &=&E_{b}\phi
_{b}\left( t\right) -K\phi _{a}\left( t\right) ,  \label{TSM_b}
\end{eqnarray}%
where
\end{subequations}
\begin{subequations}
\begin{eqnarray}
E_{a} &=&-2\kappa \sum_{\left\langle \mathbf{i},\mathbf{j}\right\rangle
}\Phi _{a,\mathbf{i}}\Phi _{a,\mathbf{j}}=-2d\kappa ,  \label{Ea} \\
E_{b} &=&\mu -2\kappa \sum_{\left\langle \mathbf{i},\mathbf{j}\right\rangle
}\Phi _{b,\mathbf{i}}\Phi _{b,\mathbf{j}}=\mu -2d\kappa ,  \label{Eb} \\
U_{a} &=&U\sum_{\mathbf{i}}\Phi _{a,\mathbf{i}}^{4}=\frac{U}{N},  \label{Ua}
\\
K &=&g\sum_{\mathbf{i}\in A}\Phi _{a,\mathbf{i}}\Phi _{b,\mathbf{i}}=\frac{g%
}{N^{\frac{1}{d}}},  \label{K}
\end{eqnarray}%
for two $d$-dimensional lattices with the periodic boundary condition and
contact surface with size $N^{\frac{d-1}{d}}$. $\phi _{a,b}\left( t\right)=%
\sqrt{N_{a,b}}e^{i\theta _{a,b}}$ where $N_{a,b}$ and $\theta _{a,b}$ are
the particle number and phase of the superfluid component in lattices A and
B.

Substituting $\phi _{a,b}\left( t\right) $ into the above two-state model,
we get
\end{subequations}
\begin{subequations}
\begin{eqnarray}
\frac{\partial z}{\partial t} &=&-2K\sqrt{1-z^{2}}\sin \Theta ,  \label{Zt}
\\
\frac{\partial \Theta }{\partial t} &=&\Delta E+\Lambda z+\frac{2Kz}{\sqrt{%
1-z^{2}}}\cos \Theta ,  \label{Theta_t}
\end{eqnarray}%
where $\Theta =\theta _{a}-\theta _{b}$, $z=(N_{b}-N_{a})/(N_{a}+N_{b})$, $%
\Delta E=E_{b}-E_{a}-\frac{U_{a}}{2}\left( N_{a}+N_{b}\right) $, and $%
\Lambda =U_{a}\left( N_{a}+N_{b}\right) /2$. Eqs. (\ref{TSM_a}) and (\ref%
{TSM_b}) show that $N_{a}+N_{b}$ is conserved. Then the Josephson current
(if the boson is neutral, the current corresponds to the flow of bosons) is
\end{subequations}
\begin{equation}
J(t)=\frac{\partial N_{b}}{\partial t}=-2K\sqrt{N_{a}N_{b}}\sin \Theta .
\label{J(t)}
\end{equation}%
When $\mu \gg U,\kappa $, it becomes
\begin{equation}
J(t)=-\frac{2g\sqrt{N_{a}N_{b}}}{N^{\frac{1}{d}}}\sin \mu t.  \label{Jt}
\end{equation}%
Here $N_{a,b}$ is time-dependent. On the other hand, the upper bound of the
particle immigration across the junction during a half period of
oscillation, which corresponds to the case $N_{a,b}\sim N$, is of the order $%
\Delta N_{b}\sim $ $g\sqrt{N_{a}N_{b}}/(\mu N^{\frac{1}{d}})$ $\sim gN^{%
\frac{d-1}{d}}/\mu $. For $N^{\frac{d-1}{d}}\ll N$, $N_{a,b}$ can be
regarded as a constant, which corresponds to the order parameter defined in
the framework of the mean field theory \cite{Fisher} as $\psi _{\gamma
}\equiv \left\langle \gamma _{\mathbf{i}}\right\rangle $ $=\left\langle
\gamma _{\mathbf{i}}^{\dag }\right\rangle $ $=\sqrt{N_{\gamma }/N}$, ($%
\gamma =a,b$). Then the Josephson current is obtained as
\begin{equation}
J(t)=-2gN^{\frac{d-1}{d}}\psi _{a}\psi _{b}\sin \mu t.  \label{J_ab}
\end{equation}

\emph{Measurement of the order parameter. }In this section, we focus on the
experimental scheme to obtain the Mott lobes of the Bose-Hubbard model. It
is based on measuring the magnitude and frequency component of the Josephson
current. Our proposal for the measurement of the order parameter
experimentally is as follows: (a) Parameters $U$, $\kappa $ and $\mu $ are
set to be values corresponding to a point ($\mu /U$, $\kappa /U$) on the
phase diagram, then cool the system to the ground state. (b) Take the ground
state as an initial state, and shift the chemical potential to $\mu +\Delta
\mu $ ($\Delta \mu \gg U,\kappa $). (c) Obtain the current $J(t)$ between A
and B. (d) Numerically analyze the curve $J(t)$ to obtain
\begin{equation}
J_{m}\left( \mu /U,\kappa /U\right) =\max \left\{ J\left( \mu /U,\kappa
/U,t\right) \right\}  \label{Jm}
\end{equation}%
and the frequency component of the current from the formula
\begin{equation}
J\left( \mu /U,\kappa /U,\omega \right) =\sqrt{\frac{2}{\pi }}\int_{0}^{\tau
}dtJ\left( \mu /U,\kappa /U,t\right) \sin \omega t  \label{Jomiga}
\end{equation}%
as $\tau \rightarrow \infty $. (e) In the weak coupling limit $g\ll \kappa $%
, all the bosons in lattice B are in the superfluid state, i.e., $\psi
_{b}\simeq \langle b_{\mathbf{i}}^{\dag }b_{\mathbf{i}}\rangle ^{1/2}$. Then
$\psi _{b}$\ can be measured via the measurement of the average particle
density in lattice B. (f) Construct the phase diagram $\psi _{a}\left( \mu
/U,\kappa /U\right) $ according to Eq. (\ref{J_ab}).

Theoretically, the Josephson current arises from the time evolution of the
superfluid state of the Hamiltonian (\ref{H1}). In this Hamiltonian, we add
the chemical potential $\mu $ only in system B, but this is equivalent to
adding a chemical potential $-\mu $ in system A, since adding a term $-\mu
\sum_{\mathbf{i}}\left( a_{\mathbf{i}}^{\dag }a_{\mathbf{i}}+b_{\mathbf{i}%
}^{\dag }b_{\mathbf{i}}\right) $ to the whole system does not bring any
physical change. Then the order parameter of lattice A corresponds to the
point ($\mu /U$, $\kappa /U$) of the obtained phase diagram. At $t=0$, the
ground state of the Hamiltonian (\ref{H1}) is $\left\vert \varphi _{g}\left(
\mu /U,\kappa /U\right) \right\rangle =\left\vert \varphi \left( 0\right)
\right\rangle $, which is the initial state for the time evolution driven by
the Hamiltonian $H+\Delta \mu \sum_{\mathbf{i}}b_{\mathbf{i}}^{\dag }b_{%
\mathbf{i}}$, i.e., $\left\vert \varphi \left( t\right) \right\rangle
=e^{-i(H+\Delta \mu \sum_{\mathbf{i}}b_{\mathbf{i}}^{\dag }b_{\mathbf{i}%
})}\left\vert \varphi \left( 0\right) \right\rangle $. Then the current
across A and B is
\begin{eqnarray}
&&J\left( \mu /U,\kappa /U,t\right)  \label{J(t)_Phi(t)} \\
&=&-ig\sum_{\mathbf{i}\in C}\left\langle \varphi \left( 0\right) \right\vert
\left( a_{\mathbf{i}}^{\dagger }\left( t\right) b_{\mathbf{i}}\left(
t\right) -h.c.\right) \left\vert \varphi \left( 0\right) \right\rangle ,
\notag
\end{eqnarray}%
where $a_{\mathbf{i}}^{\dagger }\left( t\right) $ and $b_{\mathbf{i}}\left(
t\right) $ are boson operators in the Heisenberg picture. To demonstrate
this scheme, the numerical simulation is performed for $N=2$ system with $%
\kappa =1$, $\Delta \mu =100$, $g=0.1$, and $\mu /U=0.5$, which corresponds
to a parallel line in the phase diagram of lattice A. Fig. 2(a) is the 3D
plot of the Josephson current $J\left( U,t\right) $, which is a good
sinusoidal curve vs time $t$ when $U$ is not very large, in which region
systems A and B are both in the superfluid phase. When $U$ is large enough,
the amplitude of the current drops, which indicates that system A enters
into the Mott insulating phase. Fig. 2(b) is the 3D plot of the Fourier
component $J\left( U,\omega \right) $ obtained according to Eq. (\ref{Jomiga}%
). The upper limit of the time integration (\ref{Jomiga}) is taken as $\tau
\left( 1/\Delta \mu \right) =20$. The larger values of $\tau $ will make the
peak with frequency $100$ higher and sharper. The current drops rapidly as $%
U $ increases, which indicates the quantum phase transition of system A.
These results show that the current is a Josephson alternating current with
frequency $\Delta \mu $ and can be employed to witness the quantum phase
transition.


\begin{figure}[tbp]
\includegraphics[bb=9 135 594 691, width=4 cm]{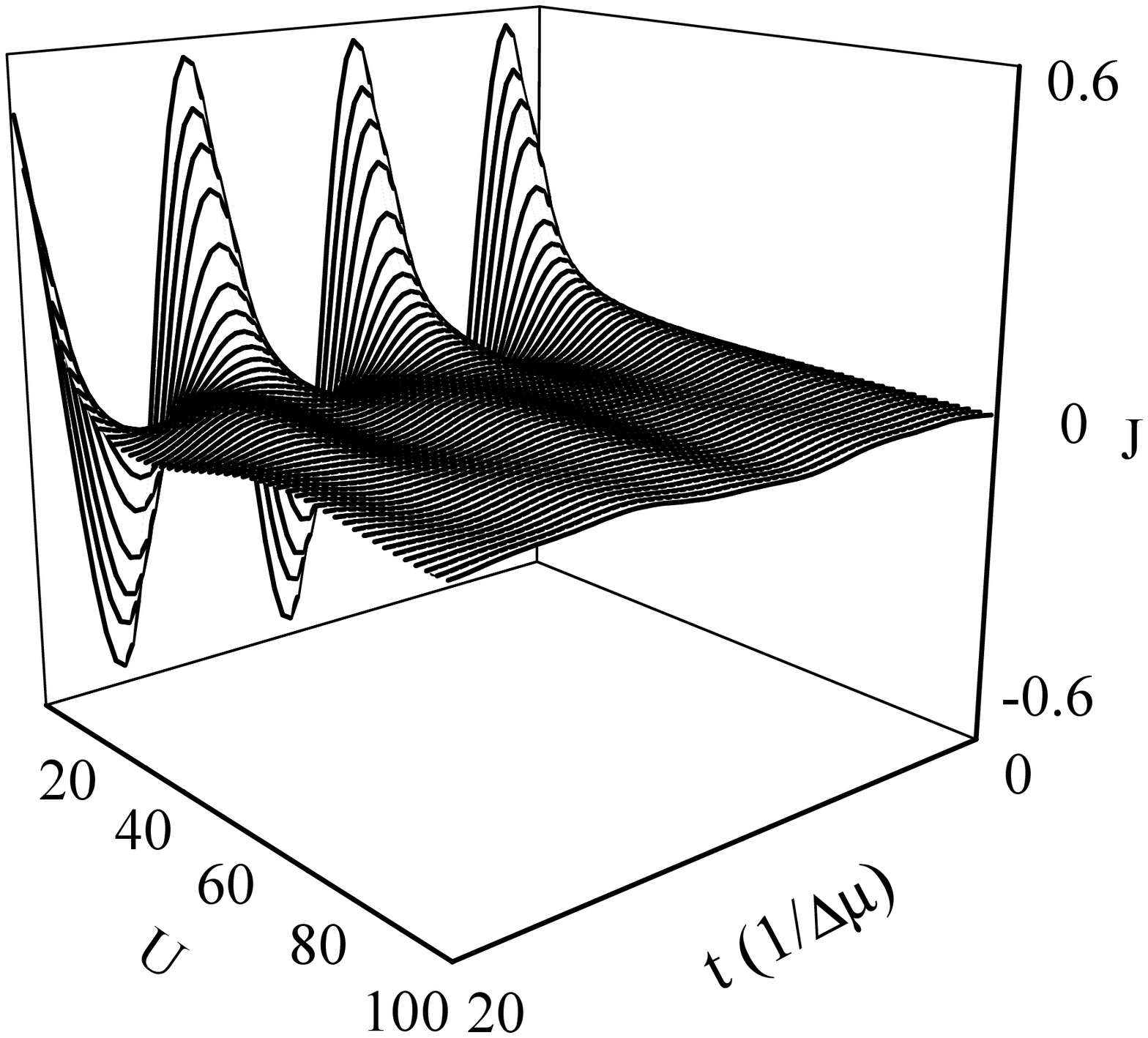} %
\includegraphics[bb=6 143 595 710, width=4 cm]{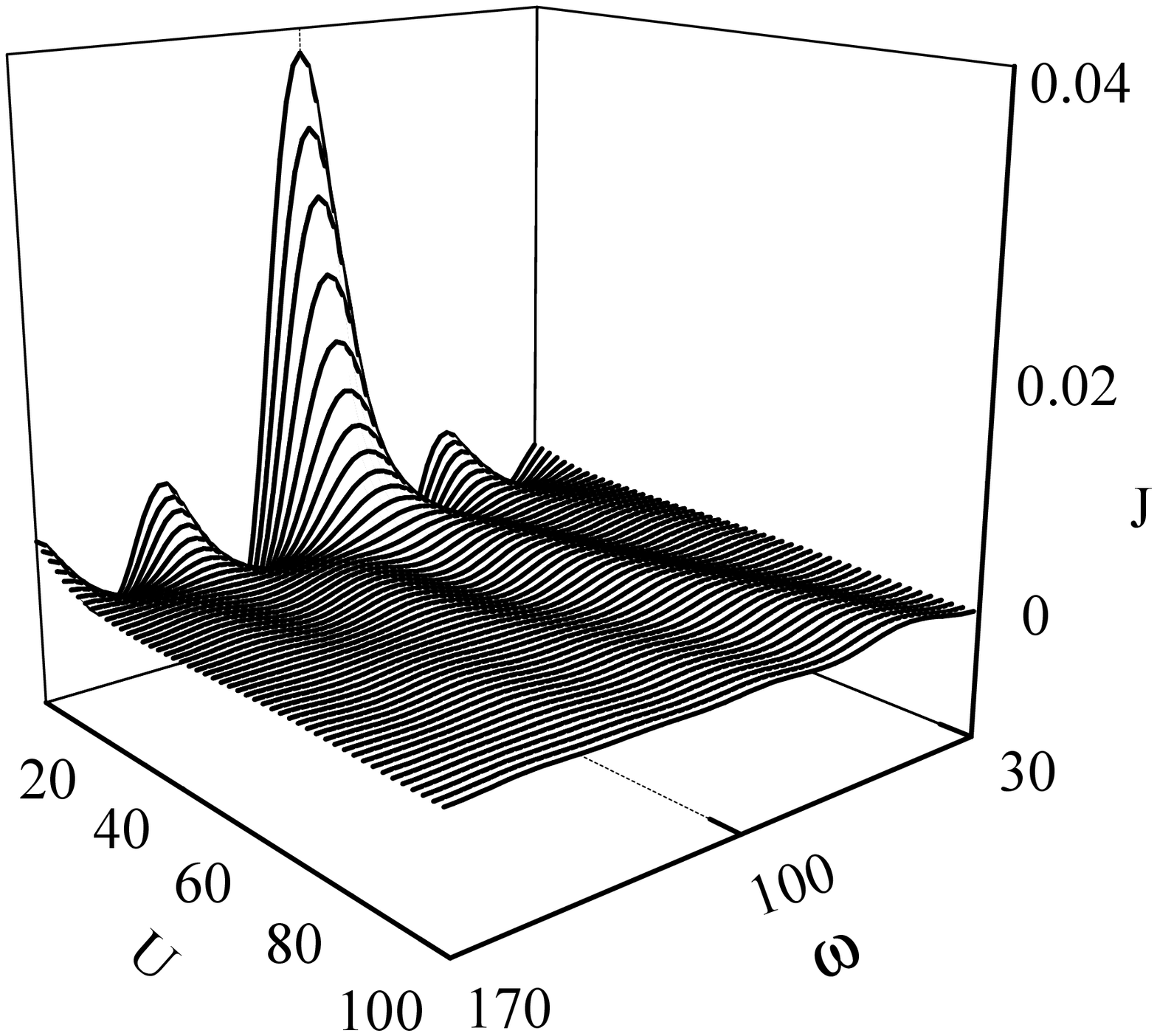}
\caption{\textit{(Color online) 3D plots of the Josephson current (a) }$%
J\left( U,t\right) $\textit{\ and its Fourier component (b) }$J\left( U,%
\protect\omega \right) $\textit{\ for a small size system obtained by the
exact diagonalization with }$\protect\kappa =1$\textit{, }$\Delta \protect%
\mu =100$\textit{, }$g=0.1$\textit{\ and }$\protect\mu /U=0.5$\textit{. In
(b), the upper limit of the time integration is taken as }$\protect\tau %
(1/\Delta \protect\mu )=20$\textit{. As }$\protect\tau $\textit{\ increases,
the peak with frequency }$100$\textit{\ will become higher and sharper. The
current drops rapidly as }$U$\textit{\ increases. These results show that
the current is a Josephson alternating current with frequency }$\Delta
\protect\mu $\textit{\ and can be employed to witness the quantum phase
transition.}}
\end{figure}


According to the mean field results, when $g$ is small enough, the Josephson
current is proportional to the order parameter of an isolated system with
the Hamiltonian $H_{A}\rightarrow H_{A}-\mu \sum_{\mathbf{i}}a_{\mathbf{i}%
}^{\dag }a_{\mathbf{i}}$. However, the Mott lobes obtained in a finite size
system from the above procedures (a-f) cannot be compared with the phase
diagram obtained from the mean field method in the thermodynamic limit,
which is usually obtained via the Gutzwiller trial wave function \cite%
{Fisher}. To compare the Josephson current with the mean field phase
diagram, we need to develop another way to obtain the order parameter for a
finite system in the framework of the spontaneous symmetry breaking. It is
well known that the quantum phase transition is a consequence of U(1)
symmetry breaking for an interacting boson system \cite{Sachdev}. Then
introducing an auxiliary field as
\begin{equation}
H_{A}-\mu \sum_{\mathbf{i}}a_{\mathbf{i}}^{\dag }a_{\mathbf{i}}\rightarrow
H_{A}-\mu \sum_{\mathbf{i}}a_{\mathbf{i}}^{\dag }a_{\mathbf{i}}+\lambda
\sum_{\mathbf{i}}\left( a_{\mathbf{i}}^{\dagger }+a_{\mathbf{i}}\right)
\label{auxiliary}
\end{equation}%
breaks the U(1) symmetry breaking and induces the order parameter.
Accordingly, the order parameter is determined by $\psi _{a}=\lim_{\lambda
\rightarrow 0}\lim_{N\rightarrow \infty }\left\langle a_{\mathbf{i}%
}\right\rangle ,$ where the average is taken for the ground state of the
Hamiltonian (\ref{auxiliary}). In the thermodynamic limit $N\rightarrow
\infty $, the ground state $\left\vert \varphi _{g}^{a}\right\rangle $ is a
coherent state and the vanishing $\psi _{a}$ discriminates two phases. In
this case, the order parameter is independent of $\lambda $. On the other
hand, when lattice A is coupled to B, the order parameter obtained from Eqs.
(\ref{J_ab}) and (\ref{J(t)_Phi(t)}) should also be independent of $g$ in
the limit $N\rightarrow \infty $ and $g\rightarrow 0$.


\begin{figure}[tbp]
\includegraphics[bb=8 347 589 784, width=7 cm]{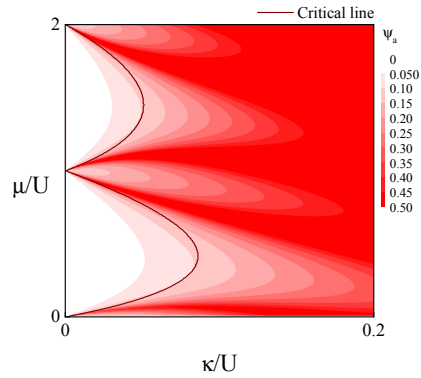}
\caption{\textit{(Color online) Mott lobes calculated by the exact
diagonalization with an auxiliary field for a small size system (color
contour map), and by the mean field method via the Gutzwiller trial wave
function. The contour line denotes the boundary of two phases which
corresponds to a vanishing order parameter. It is shown that although the
result of the small system can not give the boundary of two phases, its
contour lines are consistent with that from the mean field method in the
thermodynamic limit very well.}}
\end{figure}


In a finite system, although $\psi _{a}$ obtained from the above two ways
are not independent of $\lambda $ and $g$, it is believed that their
consistency should be revealed from the contour maps. To demonstrate this,
the numerical simulation for a small size system is performed and compared
with the mean field method. Even for a small size system, the dimension of
the Hilbert space of the Hamiltonian (\ref{auxiliary}) is infinite. So the
truncation approximation is taken for the exact diagonalization of the
matrix for $N=2$ and the average density $\left\langle a_{\mathbf{i}}^{\dag
}a_{\mathbf{i}}\right\rangle \in \lbrack 0,2]$. In Fig. 3, the Mott lobes
calculated by the exact diagonalization with an auxiliary field for $N=2$
system and by the mean field method via the Gutzwiller trial wave function
are plotted. The contour line denotes the boundary of two phases, which
corresponds to a vanishing order parameter. It is shown that although the
result of the small size system can not give the boundary of two phases, its
contour lines are consistent with that of the mean field method in the
thermodynamic limit very well.


\begin{figure}[tbp]
\includegraphics[bb=17 369 523 789, width=7 cm]{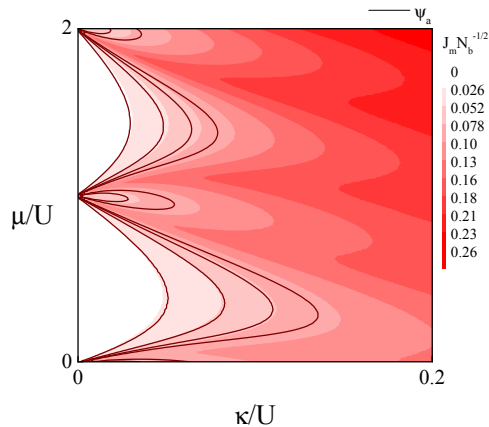}
\caption{\textit{(Color online) Mott lobes of the order parameter for a
small size system calculated by the exact diagonalization with }$\protect%
\kappa =1$\textit{, }$\Delta \protect\mu =100$, and $g=0.1$. \textit{The
color contour map is obtained from the Josephson current, while the contour
lines are obtained via an auxiliary field with }$\protect\lambda =0.1$%
\textit{. It is shown that two results are consistent very well.}}
\end{figure}


In Fig. 4, Mott lobes of the order parameter for a small size system are
plotted through two different mechanisms with $\kappa =1$ and $g=0.1$. The
color contour map is obtained from the Josephson current via the above
procedures (a-f), while the contour lines are obtained from the Hamiltonian (%
\ref{auxiliary}) by the exact diagonalization. It is shown that two results
are consistent very well, which indicates that the property of the small
size system can shed light on the profile of Mott lobes in the
thermodynamics limit.

\emph{Discussion. }In order to detect the Mott lobes of the quantum phase
transition, we investigate the AC Josephson effect in a system consisting of
two weakly coupled Bose-Hubbard models. The mean field theory in the
thermodynamic limit and the numerical simulation for a small size system
show that, through measuring the magnitude and frequency component of the
Josephson current, the Mott lobes can be measured.

To realize this scheme experimentally, as mentioned before, a good candidate
is the coupled cavity system with each cavity interacting with 4-level atoms
driven by an external laser \cite{Photonnonlinearity}. In this system, the
repulsion $U$ can indeed be strong enough to observe the Mott insulator
state for photons. Moreover, according to Refs. \cite{Photonnonlinearity}
and \cite{parameters}, the controllable range of the chemical potential
required by our scheme is in experimentally accessible parameter regimes. In
fact, in such an effective Bose-Hubbard system, the on-site interaction $U$
and chemical potential $\mu $ for photons are determined by%
\begin{equation}
U=S\left( \frac{g_{13}}{\Omega }\right) ^{2}\frac{g_{24}^{2}}{\Delta },\text{
}\mu =S\left( \frac{g_{13}}{\Omega }\right) ^{2}\epsilon ,  \label{U}
\end{equation}%
under the conditions $g,\Delta ,g_{24},\epsilon \ll \Omega $; $g_{24}g\ll
\left\vert \Delta \Omega \right\vert $. Here $\Omega $ is the Rabi frequency
of the driving laser; $S$ is the atom number in each cavity; $g_{13}$, $%
g_{24}$ are the couplings between cavity mode to the atomic levels; $\delta $%
, $\Delta $ and $\epsilon $ are detunings of atomic transitions with respect
to the cavity and laser fields. All the notations are identical with those
used in Fig. 1 of Ref. \cite{Photonnonlinearity}.

By fixing $S\left( g_{13}/\Omega \right) ^{2}$\ and adjusting $%
g_{24}^{2}/\Delta $ and $\epsilon $, $U$ and $\mu $ are tunable
independently. This allows us to simulate the Hamiltonian (\ref{Habc}) by
setting $\epsilon =0$ in lattice A to get $\mu =0$, and $g_{24}=0$ in
lattice B to get the vanishing $U$. Subsequently, to drive the quantum phase
transition and probe the Mott lobes of lattice A, the parameters in two
systems A and B are tuned according to the procedures (a-f). In this scheme,
the Josephson current $J(t)$ is the flow of photons in photonic crystal
waveguides, which can be imaged via a high-resolution imaging technique,\
the collection scanning near-field optical microscope \cite{SNOM}. This
predicts that the Mott lobes for a small quantum device can be detected
experimentally.

This work is supported by the NSFC with grant Nos. 90203018, 10474104 and
60433050, and NFRPC with Nos. 2006CB921206 and 2005CB724508.

\end{document}